\documentclass[sigconf]{aamas} 

\usepackage{balance} 
\usepackage[linesnumbered,ruled,vlined]{algorithm2e}
\usepackage{amsmath}
\usepackage{algorithmic}
\usepackage{url}

\setcopyright{ifaamas}
\acmConference[AAMAS '23]{Proc.\@ of the 22nd International Conference
on Autonomous Agents and Multiagent Systems (AAMAS 2023)}{May 29 -- June 2, 2023}
{London, United Kingdom}{A.~Ricci, W.~Yeoh, N.~Agmon, B.~An (eds.)}
\copyrightyear{2023}
\acmYear{2023}
\acmDOI{}
\acmPrice{}
\acmISBN{}

\title{A Decentralized Multiagent-Based Task Scheduling Framework for Handling Uncertain Events in Fog Computing}

\author{Yikun Yang}
\affiliation{
  \institution{University of Wollongong}
  \city{Wollongong}
  \country{Australia}}
\email{yikun@uow.edu.au}

\author{Fenghui Ren}
\affiliation{
  \institution{University of Wollongong}
  \city{Wollongong}
  \country{Australia}}
\email{fren@uow.edu.au}

\author{Minjie Zhang}
\affiliation{
  \institution{University of Wollongong}
  \city{Wollongong}
  \country{Australia}}
\email{minjie@uow.edu.au}

\begin{abstract}
Fog computing has become an attractive research topic in recent years. As an extension of the cloud, fog computing provides computing resources for Internet of Things (IoT) applications through communicative fog nodes located at the network edge. Fog nodes assist cloud services in handling real-time and mobile applications by bringing the processing capability to where the data is generated. However, the introduction of fog nodes can increase scheduling openness and uncertainty. The scheduling issues in fog computing need to consider the geography, load balancing, and network latency between IoT devices, fog nodes, as well as the parent cloud. Besides, the scheduling methods also need to deal with the occurrence of uncertain events in real-time so as to ensure service reliability. This paper proposes an agent-based framework with a decentralized structure to construct the architecture of fog computing, while three agent-based algorithms are proposed to implement the scheduling, load balance, and rescheduling processes. The proposed framework is implemented by JADE and evaluated on the iFogSim toolkit. Experimental results show that the proposed scheduling framework can adaptively schedule tasks and resources for different service requests in fog computing and can also improve the task success rate when uncertain events occur. 
\end{abstract}

         
\newcommand{\BibTeX}{\rm B\kern-.05em{\sc i\kern-.025em b}\kern-.08em\TeX}

\begin{document}


\pagestyle{fancy}
\fancyhead{}


\maketitle 

\section{Introduction}

Internet of Things (IoT) devices with embedded sensors and actuators have been widely applied in our daily life from personal use to industrial productions \cite{abd2021advanced,fellir2020multi,hossain2021scheduling}. The extensive data streams generated by IoT devices require powerful computational resources to process \cite{fellir2020multi,hossain2021scheduling}. Nowadays, cloud computing delivers powerful computing resources for IoT applications through virtual technology in a centralized manner \cite{amini2020task,talmale2021dynamic,tychalas2020scheduling}. However, the cloud centres are usually far away from end-users which involves a different degree of latency across the world. Based on the distance from the gateway to cloud centres, the latency varies from single digits millisecond range (e.g., 8 ms for AWS-Sydney) to hundreds of milliseconds (e.g., 281 ms for AWS London) \cite{latencyinfo}. Besides, due to network congestion, bandwidth obstacle, man-in-middle attacks, etc., centralized provision of computing resources is insufficient for time-sensitive tasks (e.g., healthcare critical tasks) \cite{amini2020task}.

Fog computing is the evolutionary extension of cloud computing, which attempts to bring computing resources to where the data is generated instead of moving the data to cloud centres, so as to relieve the communication amount and latency \cite{lea2018internet}. Cisco defined a fog node as any device that extends the cloud computing near end-users with storage, computing capability, and network connectivity (e.g., the Internet Service Provider (ISP) gateway) \cite{rahmani2018exploiting}. Fog computing employs fog nodes to deliver computing services and shares the API and communication standard with active fog nodes. The fog topologies are based on the hierarchy of networks, the number of IoT devices, as well as the fog node capabilities. Under the fog computing architecture, fog nodes and the cloud can communicate and migrate data through the network, thereby balancing the resource load and developing the system reliability \cite{abd2021advanced}. Nowadays, fog computing has been applied in several real-time distributed scenarios, such as smart traffic management systems, wireless sensor networks, and healthcare IoT systems \cite{stojmenovic2014fog,mutlag2019enabling}.

Task scheduling in fog computing refers to (\textbf{i}) configure the required computing resources in nodes, (\textbf{ii}) transfer task data from IoT devices to nodes, (\textbf{iii}) execute tasks in order, and (\textbf{iv}) return execution results to IoT devices. Online computing services are affected by many practical constraints, such as the geographical distances between IoT devices and nodes affecting data transmission latency. In contrast, the capabilities of nodes determine the execution speed of tasks. Scheduling methods need to take these practical constraints into account. Besides, compared with scheduling issues in conventional cloud computing, the introduction of fog nodes increases the scheduling scale, openness, and uncertainties \cite{chandak2020multi,guevara2021task,mokni2021cooperative}. 
Three challenging issues need to be considered when processing task scheduling in the context of fog computing.

\begin{description}

\item[1.] Considering the heterogeneity and wide geographic distribution of IoT devices and fog nodes, it is challenging to schedule global tasks on suitable computing resources with multiple real-world constraints in a centralized manner. Traditional centralized scheduling methods (e.g., the genetic scheduling algorithm \cite{velliangiri2021hybrid} and dispatching rules \cite{dhurasevic2018survey}) requiring global information in advance are difficult to be applied in fog computing.

\item[2.] One feature of fog computing is to load balance through communications and data migrations between fog nodes and the cloud. In geographically distributed fog computing environments, due to the dynamic change of node load and the arrival of streams of tasks, it is challenging to autonomously make decisions on data migrations and enable nodes to reach agreements on data migrations quickly.

\item[3.] Since fog nodes are located at the edge of the network without central management, fog nodes are not as stable as clouds. Additionally, IoT users may not use computing resources as planned. In this case, both the improper operations of IoT users and the instability of fog nodes can increase the scheduling uncertainties, where unexpected events (such as task cancellations and hardware damages) may appear during the scheduling process. It is challenging to deal with uncertain events with high efficiency and success rate because they are hard to detect, and decision-making on rescheduling in dynamic environments is time-cost.

\end{description}

This paper proposes an agent-based framework for task scheduling and rescheduling in fog computing to address the challenges mentioned above. In fog computing environments, agents are easy to implement, because entities in fog computing, whether the IoT devices or fog nodes, can communicate in some manner (e.g., wired or wireless, near-range or far-field) and have basic computing capabilities (e.g., embedded microprocessors). Agent technologies have demonstrated advantages in modelling complex scheduling environments, where intelligent agents can autonomously make rational decisions for task scheduling and respond to environmental changes timely \cite{gowri2019agent,mokni2021cooperative,niu2020gmas,talmale2021dynamic}. 

The three main contributions of the proposed framework are as follows. (1) The proposed framework employs intelligent agents to reconstruct the fog computing environment, where agents can access the real-time information of IoT devices and fog/cloud nodes. Besides, agents can adaptively make rational decisions for task (re-)scheduling and reach agreements on data migrations in a decentralized manner. 
(2) To fit the actual fog computing environment, the proposed framework not only considers the essential scheduling attributes, but also considers the geographic locations and load balancing. In this case, a novel agent-based algorithm for the initial task scheduling is proposed which aims to (i) minimize the completion time of tasks, (ii) balance loads of local fog nodes, and (iii) minimize the conflicts among decentralized agents caused by their self-interests. 
(3) This paper considers four types of typical uncertain events in fog computing environments. An agent-based rescheduling algorithm is proposed for (i) minimizing the impact of uncertain events and (ii) maximizing the task success rate, thereby improving the robustness and reliability of fog computing.

The rest of this paper is organized as follows. Section \ref{fog_cloud} formally defines the fog computing environment and four types of typical uncertain events. Section \ref{algorithm} gives the detailed framework designs for task scheduling and rescheduling. Section \ref{experiment} presents the experimental settings and analyses the results. In Section \ref{related_work}, related work is described. The conclusion is given in Section \ref{conclusion}.

\section{Problem Formulation}\label{fog_cloud}

This section introduces the fog computing environment and four types of uncertain events. Then the scheduling and rescheduling objectives are described.

\subsection{Fog Computing Environment}

The design of fog computing varies from simple to complex with multiple real-world constraints (e.g., data volume, the number of IoT devices, and fog node capabilities) \cite{lea2018internet}. The most straightforward fog computing paradigm could only include one fog node (e.g., a gateway), which delivers the cloud services to a set of connected sensors and actuators. In the commercial and industrial domains, the fog computing paradigm grows in complexity, including hierarchical fog nodes with different processing capabilities, where multiple cloud providers participate.

\begin{figure}[t!]
\centering
\includegraphics[width=8.5cm]{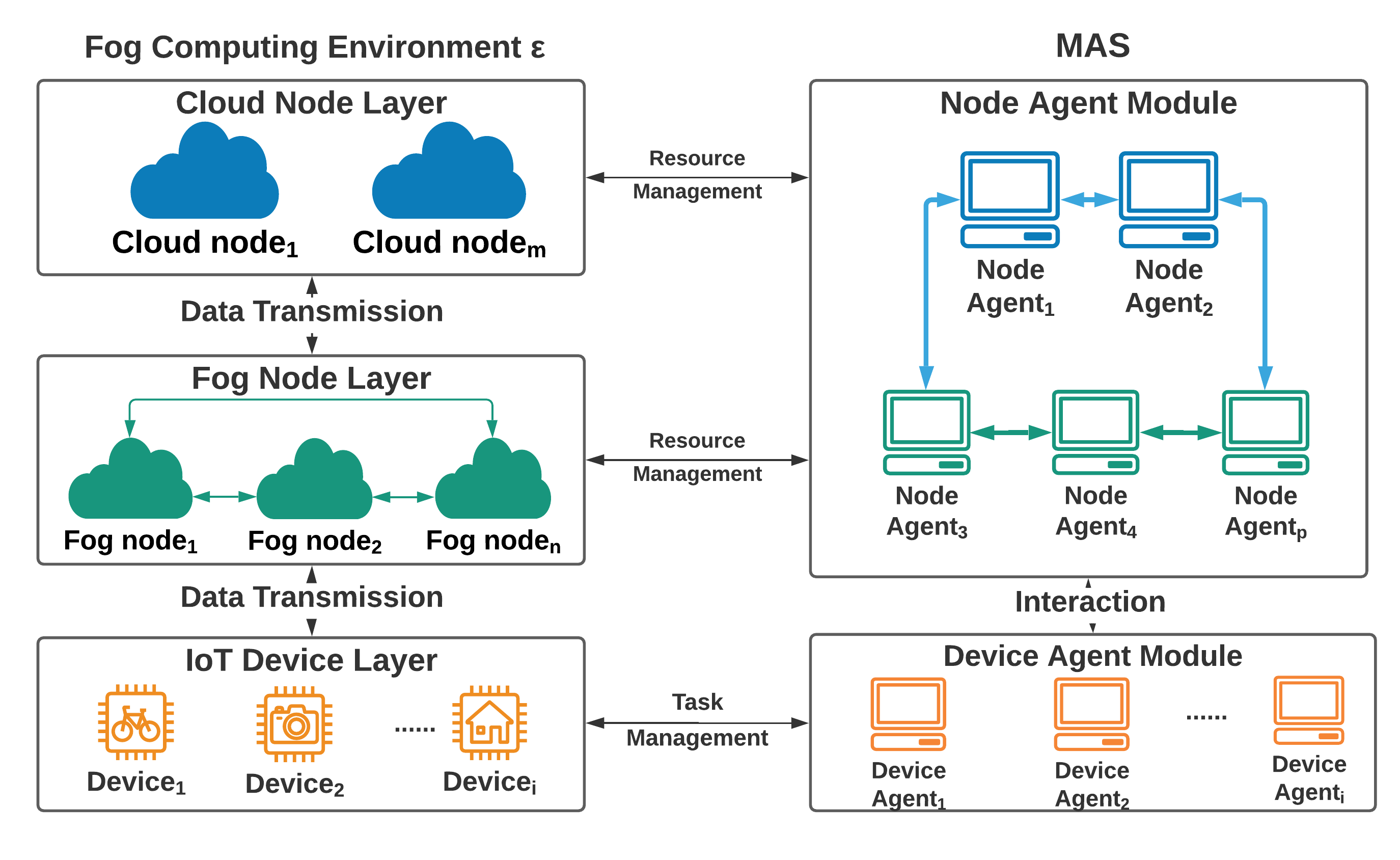}
\caption{The Architecture of the Fog Computing Environment}
\label{architecture_diagram}
\end{figure}

\textbf{Figure \ref{architecture_diagram}} shows the architecture of the fog computing environment considered in this paper and the proposed agent-based scheduling framework. This paper considers the \textit{fog computing environment} with three layers $\epsilon=\left\lbrace D, F, C  \right\rbrace$, where $D$ denotes the \textit{IoT device layer}, $F$ denotes the \textit{fog node layer}, and $C$ denotes the \textit{cloud node layer}.

The \textit{IoT device layer} $D = \left\lbrace d_{1},..., d_{i}  \right\rbrace$ includes a set of IoT devices with embedded sensors and actuators. Each IoT device $d_{i}= \left\lbrace f_{i}, gl_{i}, T_{i} \right\rbrace$ directly connects to a fog node $f_{i}$, which has a geographical location given in a two dimensional matrix $gl_{i}=\left[x_{i},y_{i}  \right]$, and includes a series of independent tasks $T_{i}=\left\lbrace t_{i}^{1},..., t_{i}^{k} \right\rbrace$ that require online computing resources to execute. A task $t_{i}^{k}$ is formally defined as Definition \ref{task}.

\begin{definition} \label{task}
The k-th task of device $d_{i}$ is defined as $t_{i}^{k}=\left\lbrace r_{i}^{k}, s_{i}^{k}, l_{i}^{k}, dl_{i}^{k} \right\rbrace$, where $r_{i}^{k}$ denotes the RAM requirement of the task, $s_{i}^{k}$ represents the size of data transmission between the $d_{i}$ and nodes, $l_{i}^{k}$ denotes the CPU length of $t_{i}^{k}$ defined as million instructions (MI), and $dl_{i}^{k}$ denotes the deadline of the task. 
\end{definition}

As shown in \textbf{Figure \ref{architecture_diagram}}, the series of tasks generated by IoT devices (i.e., the sensor components) will be sent to fog nodes or cloud nodes to be executed using online computing resources. After task executions, the computation results will be sent back to IoT devices (i.e., the actuator components). To execute the tasks, cloud nodes or fog nodes must configure the virtual computational environments in advance. Hence, to minimize the cost of data transmissions and computational configurations, this paper assumes that the series of tasks generated from the same IoT device will be allocated to the same cloud or fog node. 

The \textit{fog layer} includes a set of fog nodes $F=\left\lbrace fn_{1},..., fn_{n} \right\rbrace$, and the \textit{cloud layer} includes a set of cloud nodes $C=\left\lbrace cn_{1},..., cn_{m} \right\rbrace$. Each node owns different computing capabilities (e.g., CPU and RAM) to process tasks from IoT devices, and the network capability of each node is also different. Since fog and cloud nodes have similar attributes, let $N=F \cup C = \left\lbrace \eta_{1},..., \eta_{p} |p=m+n \right\rbrace$ denotes the set of cloud and fog nodes. A node $\eta_{p}$ is formally defined as Definition \ref{node_d}.

\begin{definition}\label{node_d}
A node is defined as $\eta_{p}=\left\lbrace gl_{p}, ram_{p}, bw_{p}, cpu_{p} \right\rbrace$, where $gl_{p}=\left[x_{p},y_{p}  \right]$ denotes the geographic location of the node, $ram_{p}$ denotes the RAM size, $bw_{p}$ denotes the bandwidth connected to $\eta_{p}$, and $cpu_{p}$ denotes the CPU processing rate given in Million Instructions Per Second (MIPS). 
\end{definition}

All nodes are directly or indirectly connected through the network. Communication between two nodes that are not directly connected will result in a waste of network resources. The network topology can be represented by an undirected graph via a $p\times p$ adjacency matrix $A_{p,p}$. In this paper, the network topology is blackboard information that can be accessed and modified by all agents in real time.

\begin{equation*}
A_{p,p} = 
\begin{pmatrix}
a_{1,1} & a_{1,2} & \cdots & a_{1,p} \\
a_{2,1} & a_{2,2} & \cdots & a_{2,p} \\
\vdots  & \vdots  & a_{i,j} & \vdots  \\
a_{p,1} & a_{p,2} & \cdots & a_{p,p} 
\end{pmatrix}
\end{equation*}
where $a_{i,j}\in \left[0,1\right]$. Here, $a_{i,j}=0$ means there is no direct network connection between $\eta_{i}$ and $\eta_{j}$, and $a_{i,j}=1$ means $\eta_{i}$ and $\eta_{j}$ is directly connected. Since the undirected graph is symmetric, so $a_{i,j}=a_{j,i}$.

\subsection{Formulation of Uncertain Events} \label{ue}

In fog computing, uncertain events such as improper operations of IoT users, hardware damages in nodes, and network congestions are unpredictable. These uncertain events can decrease the task success rate and degrade the quality of services. Let $E=\left\lbrace e_{1}, ..., e_{j} \right\rbrace$ denotes the considered uncertain events, the proposed MAS mainly focuses on solving the following four types of uncertain events.

\textbf{Type 1. IoT device changes}: An IoT device $d_{i}$ may change its task requirements before the completion. One or more parameters in a task $t_{i}^{k}$ may change, where $t_{i}^{k}\rightarrow t_{i}^{k'}$. New $t_{i}^{k'} = \left\lbrace r_{i}^{k'}, s_{i}^{k'}, l_{i}^{k'}, dl_{i}^{k'} \right\rbrace$ could consist of new RAM requirement, data size, CPU length, and deadline. For a mobile IoT device, its geographical location $gl_{i} \rightarrow gl_{i}'$ can also change, and it may connect to a new node $f_{i} \rightarrow f_{i}'$ due to the location and IP address changes.

\textbf{Type 2. Task cancellations}: An IoT device $d_{i}$ may cancel one or more tasks before completion. Hence, a task $t_{i}^{k}$ generated by $d_{i}$ can be cancelled before its completion.

\textbf{Type 3. Node capability changes}: During the run time, the capabilities of a node $\eta_{p}$ may change. The changed node $\eta_{p}'$ could consist of new RAM size $ram_{p}'$, bandwidth $bw_{p}'$, and CPU processing rate $cpu_{p}'$.

\textbf{Type 4. Node disconnection}: The network connected to a node $\eta_{p}$ may be congested or disconnected \cite{saxena2021osc}. In this case, the network topology will change, where $a_{*,p} \rightarrow 0$ and $a_{p,*} \rightarrow 0$.

\subsection{Scheduling and Rescheduling Objectives}
In fog computing, task scheduling refers to configuring virtual computing environments in cloud or fog nodes and allocating specific time slots to tasks to complete them before their deadlines. Rescheduling refers to reallocating computing resources for tasks when uncertain events occur where tasks cannot be executed as pre-scheduled. In this case, the affected tasks may be assigned to different nodes to maximize the success rate of tasks.

For optimal scheduling performance, in this paper, the main scheduling objective is (1) to minimize the task makespan, i.e., to minimize the total completion time of all generated tasks, and (2) to balance the load of fog nodes. In fog computing, the load balance usually excepts cloud nodes because they are powerful and contain multi-tires of virtual machines with different processing capabilities. In contrast, fog nodes are located at the edge and have limited capabilities, so load balance is necessary to increase the system throughput.
Besides, the rescheduling objective is to maximize the task success rate. Task makespan, load balance, and success rate are critical factors for the QoS of fog computing \cite{haghi2020quality}, which are formulated as follows.

\begin{description}

\item[1.] Let $C_{i}$ denote the task completion time of IoT device $d_{i}$, and $C_{max} = \max\limits_{d_{i} \in D} \left\lbrace C_{i} \right\rbrace$ denote the maximum completion time of all IoT devices. The first scheduling objective is the minimization of $C_{max}$, i.e., $\min C_{max}$. 

\item[2.] Let $\mu_{p}=TL_{p}/C_{max}^{p}$ denote the load of the node $\eta_{p}$, where $TL_{p}$ denotes the allocated time length for $\eta_{p}$, and $C_{max}^{p}$ denote the latest completion time of tasks in the schedule $Sch_{p}$. Let $V(\mu) = \sum_{p=1}^{p} (\mu_{p}-\bar{\mu})^2/n$ denote the load variance, where $n$ denotes the total number of fog nodes, and $\bar{\mu}$ denotes the average node load. The second scheduling objective is to minimize the load variance, i.e., $\min V(\mu)$.

\item[3.] Let $SR = SN/TN$ denote the task success rate, where $SN$ denotes the number of successful tasks after executions, and $TN$ denote the total number of tasks submitted by IoT devices. The rescheduling objective is to maximize the task success rate, i.e., $\max SR$. 

\end{description}

\section{Agent-based Framework}\label{algorithm}

This section introduces the architecture of the proposed agent-based scheduling framework and describes the scheduling and rescheduling processes through three algorithms.

\subsection{Framework Architecture} \label{MAS-arc}

This paper employs intelligent agents to model the scheduling framework for fog computing to facilitate the interplay between IoT devices and nodes. The advantages of agent technologies are as follows. (1) The distributed nature of agents is in line with fog computing, where agents can cooperate, coordinate, and negotiate with each other regardless of physical locations. (2) Intelligent agents can (i) access the real-time information of IoT devices and nodes, (ii) respond to real-time information changes, and (iii) make rational decisions on scheduling and rescheduling in a decentralized manner.

\textbf{Figure \ref{architecture_diagram}} shows the architecture of the proposed agent-based framework. Two agent modules are involved, including the (1) \textit{device agent module} and (2) \textit{node agent module}.

\textit{Device agent module} $DA=\left\lbrace da_{1},..., da_{i} \right\rbrace$ includes a set of device agents. Each device agent $da_{i}$ is responsible for an IoT device $d_{i}$. The i-th device agent $da_{i}$ is defined as $da_{i}=\left\langle ID_{i}, Task_{i} \right\rangle$, where $ID_{i}$ is the agent ID, and $Task_{i} = \left\lbrace t_{i}^{1},..., t_{i}^{k}  \right\rbrace$ stores the information of tasks generated by $d_{i}$.

\textit{Node agent module} $NA = \left\lbrace na_{1},..., na_{p} \right\rbrace$ includes a set of node agents. Each node agent $na_{p}$ is responsible for one node $\eta_{p}$. The p-th node agent $na_{p}$ is defined as $na_{p} = \left\langle ID_{p}, \eta_{p}, Sch_{p} \right\rangle$, where $ID_{p}$ is the agent ID, $\eta_{p}$ denotes the computing capabilities of the managed node, $Sch_{p} = \left[ d_{p}^{1},..., d_{p}^{i} \right]$ is the managed schedule that includes the expected execution time slots of IoT devices, where $d_{p}^{i} = \left[ ST_{p}^{i}, CT_{p}^{i} \right]$, $ST_{p}^{i}$ denotes the expected start time, and $CT_{p}^{i}$ denotes the expected completion time.

\subsection{The Procedure of Initial Task Scheduling}\label{task_scheduling}

\textbf{Initial task scheduling} refers to allocating computing resources in cloud or fog nodes for tasks generated by IoT devices. This section gives the agent-based algorithm for task scheduling, which describes the agent interactions and decision-making for two types of agents. The proposed algorithm takes the geographic locations of devices and nodes, as well as the network topology, into the consideration and tries to (1) minimize the makespan of tasks, (2) balance the load of fog nodes, and (3) minimize the conflicts among agents caused by their self-interests.

The initial scheduling process consists of four steps: (1) \textit{information access}, (2) \textit{node search}, (3) \textit{agent interaction}, and (4) \textit{task execution}.
Algorithm \ref{alg_1} illustrates how agents help an IoT device $d_{i}$ to use computing resources in an online node. The inputs of Algorithm \ref{alg_1} are the three-layer fog computing environment $\epsilon$, the network topology $A_{p,p}$, and two agent modules $DA$ and $NA$. The output of Algorithm \ref{alg_1} is to schedule tasks for an IoT device $d_{i}$ with specified time slots.

\begin{algorithm}[h] 
 \caption{Initial Task Scheduling for an IoT Device}
 \label{alg_1}
\SetAlgoLined
\SetKwInOut{Input}{input}
\SetKwInOut{Output}{output}   
\Input{$\epsilon=\left\lbrace D, F, C \right\rbrace$, $A_{p,p}$, $DA$, $NA$}
\Output{schedule tasks for $d_{i}$ with specified time slots}
$da_{i}$ accesses information of device $d_{i}$ in real-time\; \label{access_1}

\While{$d_{i} \in D$ generates tasks $T_{i}=\left\lbrace t_{i}^{1},...,t_{i}^{k} \right\rbrace$}{
$da_{i}$ prioritizes tasks in $Task_{i}$, the closer the $dl_{i}^{k}$, the higher the priority\; \label{access_2}
$da_{i}$ breadth-first searches for nodes based on the $A_{p,p}$, where nodes with the same depth are sorted by distance $\gamma(gl_{i}, gl_{p})$\;\label{rank_node_1}

$da_{i}$ selects $\theta \in \mathbb{N_+}$ number of node agents with high-priority to negotiate\; \label{initiate_1}

\For{each selected node agent $na_{p}$}{

$da_{i}$ sends $T_{i}$ to $na_{p}$\;\label{ana_1}

$na_{p}$ evaluates the task requirements and calculates $ET_{p}^{i}$ and searches for available $\left[ST_{p}^{i}, CT_{p}^{i} \right] \in Sch_{p}$\;\label{ana_2}
\uIf{$\eta_{p}$ satisfies $T_{i}$ $\&\&$ $\exists\left[ST_{p}^{i}, CT_{p}^{i} \right]$ satisfies $ET_{p}^{i}$ \label{meet_1}}{
$na_{p}$ replies a proposal $pro_{p}^{i}$ to $da_{i}$, $pro_{p}^{i} = \left[ST_{p}^{i}, CT_{p}^{i} \right]$\;}
\Else{
$na_{p}$ refuses $da_{i}$\; \label{meet_2}
}
}
$da_{i}$ puts received proposals in an array $P_{i}$\;\label{acc_1}
\eIf{$P_{i} \neq \emptyset$}{
$da_{i}$ sorts proposals by $CT_{p}^{i}$\;
$da_{i}$ accepts the winner proposal and rejects others\;
}
{
$da_{i}$ re-initiates interactions to subsequent $\theta$ number of node agents\; \label{acc_2}
}
the winner $na_{p}$ adds $\left[ST_{p}^{i}, CT_{p}^{i} \right]$ to $Sch_{p}$\; \label{exe_1}
$da_{i}$ submits tasks $T_{i}$ in order to node $\eta_{p}$\; \label{exe_2}
}
\end{algorithm}

\textbf{Information access}: The device agent $da_{i}$ is responsible for timely monitoring the information changes of the IoT device $d_{i}$. Once the device generates a set of tasks, the device agent stores the information of tasks in the $Task_{i}$ list and prioritizes tasks according to their deadline $dl_{i}^{k}$. The closer the deadline, the higher the priority (Lines \ref{access_1}-\ref{access_2}).

\textbf{Node search}: The device agent searches for and prioritizes nodes based on the network topology $A_{p,p}$ in breadth-first order. Firstly, the device agent $da_{i}$ treats the directly connected fog node $f_{i}\leftarrow\eta_{p}$ as root and sets the $depth = 0$. Then $da_{i}$ searches for nodes connected to the root node, whose $a_{p,*}=1$, sets their $depth+1$, and ignores nodes already travelled. The $da_{i}$ dynamically searches all nodes and sets depth, where nodes with the same depth are prioritized by the geographical distances $\gamma(gl_{i},gl_{p})$. The higher the distance $\gamma$, the lower the priority (Line \ref{rank_node_1}). 

In this paper, device agents prefer to select nodes with lower depth and closer geographical distance. This preference aims to improve the resource load in local fog nodes and minimize the cost of bandwidth and data transmission time. With a reduced transmission time, real-time applications will benefit from improved response time.

Next, $da_{i}$ initiates interactions with $\theta \in \mathbb{N_+}$ number of high-priority node agents (Line \ref{initiate_1}). The value of $\theta$ can influence the interaction efficiency and success rate of reaching agreements, and affect the scheduling optimization. The parameter $\theta$ will be discussed in \textbf{Section \ref{experiment}} in detail.

\textbf{Agent interaction}: After receiving the request from $da_{i}$, node agents will analyse the task requirements and calculate the sum of data transmission time and CPU processing time $ET_{p}^{i}$. The node agent will firstly check if its managed node meet the RAM requirements of tasks, i.e., $ram_{p}\geq \max r_{i}^{k}$. Then the $ET_{p}^{i}$ can be calculated as Equation \ref{trams_time}.

\begin{equation}\label{trams_time}
ET_{p}^{i} = \gamma(gl_{i}, gl_{p})*\dfrac{\sum^{k}_{k=1} s_{i}^{k} }{bw_{p}} + \dfrac{\sum^{k}_{k=1} l_{i}^{k}}{cpu_{p}}
\end{equation}
where $\gamma(gl_{i}, gl_{p})$ is the distance between the $d_{i}$ and $\eta_{p}$, $\sum_{k=1}^{k} s_{i}^{k}$ denotes the total data size of tasks $T_{i}$, $bw_{p}$ denotes the bandwidth of the node, $\sum_{k=1}^{k} l_{i}^{k}$ denotes the total task length, and $cpu_{p}$ denotes the CPU processing rate of the node (Lines \ref{ana_1}-\ref{ana_2}). 
Then each node agent $na_{p}$ will search for the available time slot in its schedule $Sch_{p}$. If there exists an available time slot to meet $ET_{p}^{i}$, $na_{p}$ will generate a proposal $pro_{p}^{i} = \left[ST_{p}^{i}, CT_{p}^{i} \right]$ that includes the expected task start time $ST_{p}^{i}$ and completion time $CT_{p}^{i}$, subject to $CT_{p}^{i} = ST_{p}^{i}+ET_{p}^{i}$. Otherwise, $na_{p}$ will refuse $da_{i}$ (Lines \ref{meet_1}-\ref{meet_2}).

During this period, $da_{i}$ waits for replies from node agents and stores the received proposals in an array $P_{i}$. If $da_{i}$ receives one or more proposals after waiting, $da_{i}$ will rank proposals based on the completion time $CT_{p}^{i}$. The closer the $CT_{p}^{i}$, the higher the priority. Then $da_{i}$ will accept the proposal that provides the earliest completion time and reject others. This preference aims to minimize the total completion time of tasks, i.e., the makespan. Otherwise, $da_{i}$ will initiate interactions with subsequent $\theta$ number of node agents in order until it reaches an agreement with a node agent (Lines \ref{acc_1}-\ref{acc_2}). 

\textbf{Task Execution}: After the proposal is accepted, the winner node agent $na_{p}$ will update its schedule $Sch_{p}$ based on $pro_{p}^{i}$. Next, tasks from $d_{i}$ will be executed on the node $\eta_{p}$ according to the time slot specified in the proposal, and $da_{i}$ will issue tasks according to their priorities (Lines \ref{exe_1}-\ref{exe_2}). 

\subsection{The Procedure of Load Balance}

Node agents mainly process the procedure of load balance. Algorithm \ref{alg_3} describes how node agents balance the fog load through agent interactions and decision-making. The inputs of Algorithm \ref{alg_3} are the network topology, device agents, and node agents. The output of Algorithm \ref{alg_3} is to balance the load between fog nodes. 

\begin{algorithm}[h!] 
 \caption{The Procedure of Load Balance}
 \label{alg_3}
\SetAlgoLined
\SetAlgoVlined
\SetKwInOut{Input}{Input}
\SetKwInOut{Output}{Output}      
\Input{$A_{p,p}$, $DA$, and $NA$}
\Output{to minimize the load variance $V(\mu)$}
\While{$Sch_{p}\neq\emptyset$ \label{schedule_1}}{
node agent $na_{p}$ periodically calculates $\mu_{p}$\;\label{schedule_2}
$na_{p}$ interacts and calculates the variance $V(p,*)$ with each connected node $\eta_{*}$ whose $a_{p,*}=1$\;\label{cal_V1}
\If{$V(p,*) \geq \delta$ $\&\&$ $\mu_{p}>\mu_{*}$ \label{check_v1}}{
$na_{p}$ checks the capability and schedule $Sch_{*}$ of $\eta_{*}$\;\label{decide_mig1}
$na_{p}$ searches for tasks in $Sch_{p}=\left\lbrace d_{p}^{1},...,d_{p}^{i}  \right\rbrace$ in reverse order\;\label{decide_mig2}
\If{$\exists d_{p}^{i}$ can be handled by $\eta_{*}$\label{contract_1}}{
$na_{p}$ coordinates $da_{i}$ and $na_{*}$ to set a contract\;\label{contract_2}
$\eta_{p}$ migrates the task requirements to $\eta_{*}$\;\label{migrate}
}
}
}
\end{algorithm}

Because node agents only have the local views, in this paper, node agents balance the fog load in a decentralized and ad-hoc manner. Each node only migrates data and balances the load with nearby connected nodes. This preference reduces the cost of data transmission, as well as minimizes the agent decision-making and interaction time. As shown in Algorithm \ref{alg_3}, when the schedule $Sch_{p}$ of a node $\eta_{p}$ is not empty, the node agent $na_{p}$ will periodically calculate the load $\mu_{p}$ of its node (Lines \ref{schedule_1}-\ref{schedule_2}). After $na_{p}$ figures out $\mu_{p}$ each time, it will interact with connected node agents to access their load $\mu$. Then $na_{p}$ calculates the variance $V(p,*)$ between its load and all connected nodes, whose $a_{p,*}=1$ (Line \ref{cal_V1}). If the variance is greater than the threshold $\delta$ and $\mu_{p}>\mu_{*}$, the node agent $na_{p}$ will decide to migrate some tasks to $\eta_{*}$ (Line \ref{check_v1}). First, the node agent $na_{p}$ checks the capability and available time slots of the node $\eta_{*}$. Then $na_{p}$ searches for tasks that can be handled by $\eta_{*}$ in its schedule $Sch_{p}$ in the reverse order (Lines \ref{decide_mig1}-\ref{decide_mig2}). If there is any task that can be executed by $\eta_{*}$, the node agent $na_{p}$ will coordinate the device agent $da_{i}$ and node agent $na_{*}$ to set a new contract (Lines \ref{contract_1}-\ref{contract_2}). After the contract is settled, the node $\eta_{p}$ will migrate relevant data to the new node $\eta_{*}$ (Line \ref{migrate}).

\subsection{The Procedure of Task Rescheduling}\label{task_rescheduling}

\textbf{Task rescheduling} refers to the process of re-allocating computing resources for tasks when uncertain events occur, and tasks cannot be completed as pre-scheduled. Since there are no third-party agents with global information in our framework, device agents and node agents with local views implement the task rescheduling process. Due to the lack of global information, solutions for uncertain events must be found through agent interactions. In order to reduce the impact of uncertain events on the execution of other tasks, and to reduce the communication volume among agents, in our framework, agents first try to resolve uncertain events peer-to-peer through ad-hoc communication. When uncertain events cannot be resolved in the local range, agents will try to expand the range of search until uncertain events are resolved. Since tasks considered in this paper are all independent, when uncertain events occur, agents only re-schedule tasks that cannot be completed. In contrast, unaffected tasks continue to execute according to the settled schedule.

Algorithm \ref{alg_2} illustrates the task rescheduling process. The inputs of Algorithm \ref{alg_2} are the fog computing environment $\epsilon$, network topology $A_{p,p}$, two agent modules $DA$ and $NA$, and uncertain events $E$. The output of the algorithm is to re-allocate available resources for tasks when uncertain events occur. 

\begin{algorithm}[h!] 
 \caption{Rescheduling under Uncertain Events}
 \label{alg_2}
\SetAlgoLined
\SetAlgoVlined
\SetKwInOut{Input}{Input}
\SetKwInOut{Output}{Output}      
\Input{$\epsilon = \left\lbrace D,F,C \right\rbrace$, $A_{p,p}$, $DA$, $NA$, $E=\left\lbrace e_{1}, ..., e_{j} \right\rbrace$}
\Output{re-allocate resources for tasks when $E$ occur}

\uIf{$\exists e_{j}$ in an IoT device $d_{i}$}{
\uIf{$e_{j} \in Type\,1$ \label{type_1_1}}{
\uIf{$t_{i}^{k} \rightarrow t_{i}^{k'}$}{
$da_{i}$ sends new task requirements $t_{i}^{k'}$ to $na_{p}$\;
\uIf{$\eta_{p}$ can handle $t_{i}^{k'}$}{
$da_{i}$ and $na_{p}$ sets new contract\;
}\Else{
$da_{i}$ searches for new node $\eta_{p'}$ to handle $t_{i}^{k'}$\;\label{type_1_2}
}

}\uElseIf{$gl_{i} \rightarrow gl_{i}'$\label{type_1_3}}{
$da_{i}$ re-calculates depth of nodes through $A_{p,p}$\;
\If{$da_{i}$ has contract with a node $\eta_{p}$}{
$da_{i}$ searches for new nodes and compares the $ET$\;
\If{$\exists \eta_{p}'$ with lower $ET$}{
$da_{i}$ terminates the contract with $\eta_{p}$ and sets new contract with $\eta_{p}'$\;\label{type_1_4}
}
}
}
}

\ElseIf{$e_{j} \in Type \,2$\label{type_2_1}}{
$da_{i}$ notifies $na_{p}$ the cancelled tasks\;
$na_{p}$ revises the schedule $Sch_{p}$\;\label{type_2_2}
}
}

\ElseIf{$\exists e_{j}$ in a node $\eta_{p}$}{
\uIf{$e_{j} \in Type\,3$\label{type_3_1}}{
$na_{p}$ checks tasks in $Sch_{p}$\;
\If{$\eta_{p}$ cannot complete $d_{p}^{i}$ in time}{
$na_{p}$ notifies $da_{i}$\;
$na_{p}$ revises the schedule $Sch_{p}$\;
$da_{i}$ searches for new nodes\;\label{type_3_2}
}
}\ElseIf{$e_{j} \in Type\,4$\label{type_4_1}}{
node agents who connect to $na_{p}$ can detect its disconnection, where $a_{p,*} \rightarrow 0$\;
device agents recognizes the change of $A_{p,p}$\;
\If{$da_{i}$ has unfinished tasks on $na_{p}$}{
$da_{i}$ searches for new nodes\;\label{type_4_2}
}
}
}
\end{algorithm}

As shown in Algorithm \ref{alg_2}, uncertain events can occur in IoT devices and nodes. Four types of uncertain events require different solutions. Device agents are mainly responsible for solving \textit{Type 1} and \textit{Type 2} uncertain events. For an uncertain event of \textit{Type 1}, the device agent will firstly check the changed information of the device. If the task requirements of the device were changed and the previous task requirements $t_{i}^{k}$ were already allocated to a node $\eta_{p}$, the device agent will send the new task requirements $t_{i}^{k'}$ to the node agent $na_{p}$. In this case, if the node can handle new task requirements, the node agent will set a new contract with the device agent. Otherwise, if the $\eta_{p}$ cannot satisfy the new task requirements (e.g., $\eta_{p}$ cannot satisfy the RAM requirements, or it cannot provide available time slots for tasks), the node agent will notify the device agent. Then the device agent will searches for new node that can handle $t_{i}^{k'}$ (Lines \ref{type_1_1}-\ref{type_1_2}). For a mobile IoT device, once its location changes, the directly connected node may change. In this case, the device agent will re-calculate the depth of each node and search for new nodes that can handle its task requirements. During this stage, if the device agent can find a new node that can handle its task requirements and provide a lower transmission time and processing time, the device agent will terminate the previous contract with the old node and set a new contract with the new node (Lines \ref{type_1_3}-\ref{type_1_4}). For an uncertain event of \textit{Type 2}, the device agent will notify the node agent of the task cancellations. Then the node agent will revises its managed schedule (Lines \ref{type_2_1}-\ref{type_2_2}).

Node agents are responsible for solving \textit{Type 3} and \textit{Type 4} uncertain events. If the node capability of a node $\eta_{p}$ changes (\textit{Type 3}), the node agent $na_{p}$ will check the allocated tasks in its schedule. If the new capability of $\eta_{p}$ cannot meet the task requirements from a device $d_{i}$, the node agent will notify $da_{i}$. Then the $na_{p}$ revises its schedule, while $da_{i}$ re-start to search for new nodes for its tasks (Lines \ref{type_3_1}-\ref{type_3_2}). If a node $\eta_{p}$ disconnects with other nodes (\textit{Type 4}), the node agents who connect to the $na_{p}$ will detect the disconnection. These node agents will modify the network topology, where $a_{p,*} \rightarrow 0$. Device agents will recognize the changes in network topology and monitor its task allocations. If a device $d_{i}$ has unfinished tasks on a disconnected node, the device agent $da_{i}$ will start to search for new nodes for its unfinished tasks (Lines \ref{type_4_1}-\ref{type_4_2}). 

\section{Experiments and Discussions}\label{experiment}

In the experiment, entities of fog computing were modelled by the iFogSim toolkit, including the IoT devices, fog and cloud nodes, and the network topology. The proposed agent-based framework was implemented in JADE, and a computer with an Intel(R) Core(TM) i7-7600 CPU@2.90 GHz and a 16.0 GB RAM was used. To evaluate the proposed framework, two experiments were conducted. \textbf{Experiment 1} aims to evaluate the \textit{initial task scheduling} and \textit{load balance} algorithms. \textbf{Experiment 2} aims to test the \textit{rescheduling algorithm}.

\subsection{Experiment 1: Settings and Results}

In Experiment 1, we consider the fog computing environment includes 2 cloud nodes, 48 fog nodes, and 1000 IoT devices. The settings of nodes are shown in \textbf{Table \ref{E1_Node}}. The settings of IoT devices are shown in \textbf{Table \ref{E1_device}}.

\begin{table}[t]
  \caption{Cloud and Fog Node Settings}
   \setlength{\tabcolsep}{1.9mm}
  \label{E1_Node}
  \begin{tabular}{lllll}\toprule
    \textit{Nodes} & $gl$ & $ram$ (GB) & $bw$ (Gb/s) & $cpu$ (MIPS)\\ \midrule
     Clouds & [0,0] &  16  & 10 &  50000       \\
     Fogs &  [0-500, 0-500] &  [2, 8] & [1, 5]& [5000, 10000]\\ 
   \bottomrule
  \end{tabular}
\end{table}

\begin{table}[t]
\caption{IoT device settings}
\setlength{\tabcolsep}{0.3mm}
 \label{E1_device}
\begin{tabular}{lcccccc}\toprule

\textit{Parameters}& $gl$ &$T$ size & $r$ (GB) & $s$ (GB) & $l$ (MI) & $dl$ (sec)\\
\hline
\textit{Values} & [0, 500] &[10, 25] & [0.1. 8] & [0.1, 0.2] & [1000, 2000] & [2000,3000]\\
\bottomrule
\end{tabular}
\end{table}

We tested the performance in terms of \textit{makespan}, \textit{load balance}, and \textit{network usage} \cite{fellir2020multi}. The makespan refers to the task completion time of all IoT devices, which includes agent decision-making, data transmission, and task execution time. The load balance is evaluated by the load variance $V(\mu)$ after the task completion. Network usage refers to the total number of bytes communicated by IoT devices and nodes. To verify the benefits of our proposed algorithms, a traditional method \textit{Round-Robin}, a heuristic algorithm proposed by Rehman et al. \cite{rehman2018min}, and a geo-aware agent-based approach proposed by Niu et al. \cite{niu2020gmas} were selected as the comparison models.


\begin{figure}[t!]
\centering
\includegraphics[width=8.5cm]{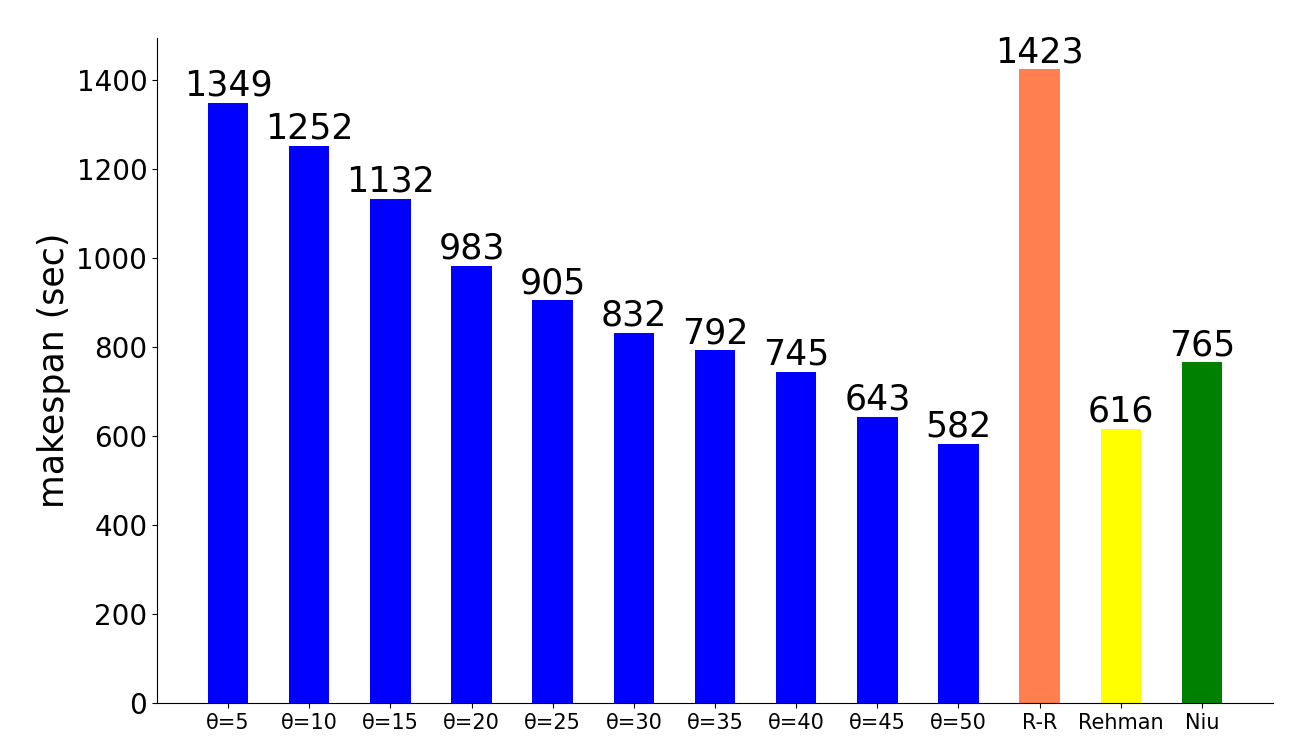}
\caption{Makespan}
\label{make_span}
\end{figure}

\begin{table}[t]
\caption{Load balance}
\setlength{\tabcolsep}{0.3mm}
 \label{l_balance}
\begin{tabular}{c|c|c|c|c}\toprule

Approaches&Our framework &R-R& Rehman et al.& Niu et al.\\
\hline
$V(\mu)$&8.954e-6& 2.835e-2&7.528e-2&4.236e-2\\
\bottomrule
\end{tabular}
\end{table}

\begin{figure}[t!]
\centering
\includegraphics[width=8.5cm]{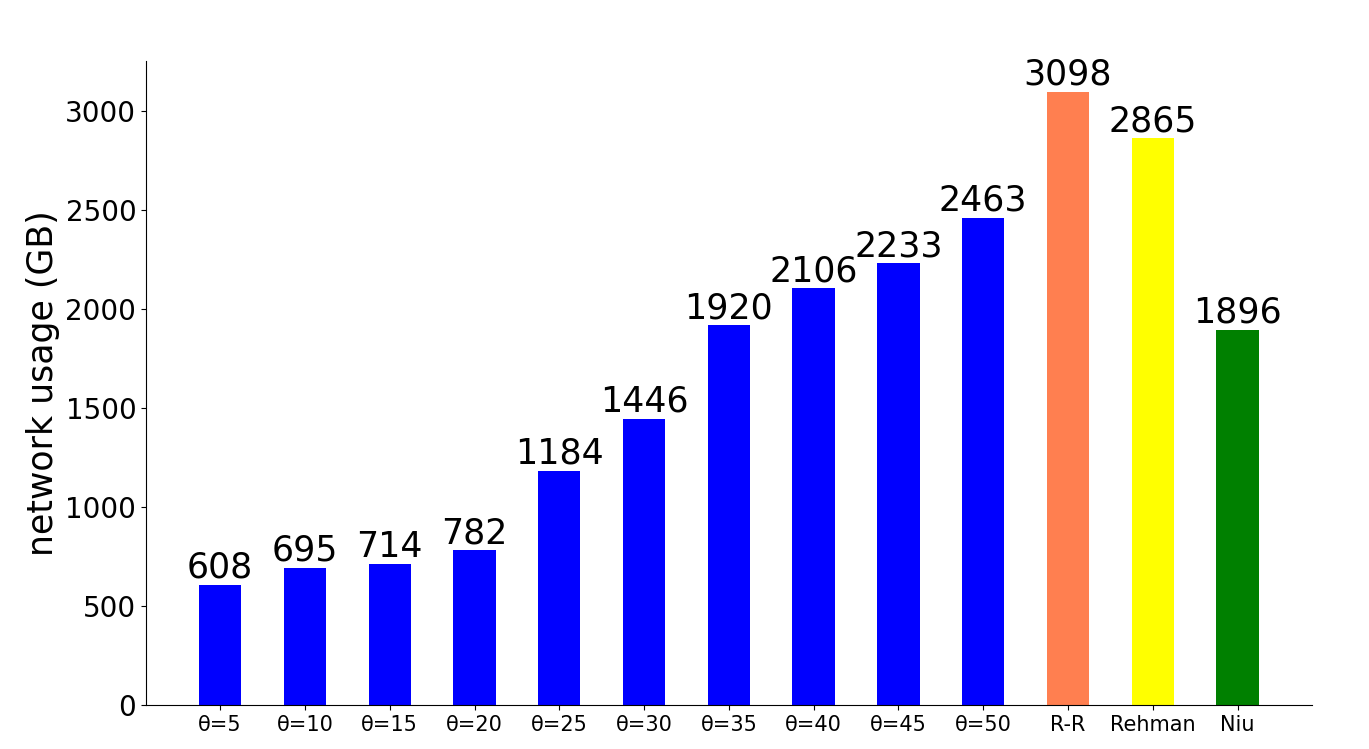}
\caption{Network usage}
\label{n_usage}
\end{figure}

\textbf{Figure \ref{make_span}} shows the measured makespan when the number of $\theta \in \left[ 10, 50 \right]$ takes different values. The value of $\theta$ represents the size of agent interactions as described in Section \ref{task_scheduling}. In addition, \textbf{Figure \ref{make_span}} also shows the experimental results of three comparison methods under the same settings. When the value of $\theta$ is small (from 10 to 30), the framework takes more than 800 seconds to complete all tasks from 1000 IoT devices. Since device agents consider the network topology and geographical distance $\gamma$ when selecting nodes, in this case, device agents only consider a few fog nodes with close distance and small depth, and two cloud nodes are almost idle. Although the close distance can reduce the response and data transmission time, the processing power of near nodes may not be powerful. When $\theta$ increases to 50, device agents consider and receive more high-quality proposals from remote nodes, thereby shortening the execution time. 

As for the three comparison methods, the Round-Robin scheme allocates resources for tasks equally in order, resulting in a long makespan (around 1423 sec). The heuristic algorithm proposed by Rehman et al. calculates the shortest execution time of tasks with global information, which can achieve an optimal makespan (around 616 sec). However, the optimal makespan of Rehman et al. method is longer than our best makespan (582 sec). This is because the method of Rehman et al. allocates resources for tasks in a centralized manner, which involves high decision-making costs. The MAS proposed by Niu et al. has a similar agent interaction protocol with our method when $\theta=50$. Their makespan (765 sec) is slightly higher than that of our framework under the same settings. Because their Geo-aware MAS takes the node distance as the main consideration, where agents only tend to choose close nodes. Our device agents select nodes not only by distance, but also by the network topology and quality of the proposal. Selecting nodes with small depths can reduce the consumption of bandwidth, so as to reduce the time on data transmissions. Hence, our method is able to balance the trade-off between the node distance and makespan optimization when $\theta$ selects different values. 

\textbf{Table \ref{l_balance}} compares the load balance between our framework and three comparison methods. \textbf{Table \ref{l_balance}} shows that our proposed load balance algorithm can reduce load variances between nodes through agent ad-hoc communications to the negative 6th power. Since the three comparison methods have no load balance considerations, they involve variances to a negative square level.

\textbf{Figure \ref{n_usage}} shows the network usage for completing tasks generated by 1000 IoT devices. As \textbf{Figure \ref{n_usage}} shows, when the $\theta$ increases, the network usage increases from 608 to 2463 GB in a near-linear tend. When the $\theta$ values are small (from 5 to 20), the device agents allocate their tasks to near nodes with small depths, so the data transmission will not be across too many nodes. The data of tasks will only be transmitted by a few nodes to reach the end. However, when the $\theta$ increases to more than 25, device agents tend to select remote nodes with more optimal completion times, where the data of tasks may need to be across too many nodes to reach the end, so as to result in a high network usage. Because the two non-agent-based comparison methods do not consider network usage, they cost a higher network usage compared to our framework. To be noteworthy, the Geo-aware agent-based method (Niu et al.) can reduce network usage to 1896 GB, because their approach takes the geographical distances between devices and nodes as the main consideration. 

\subsection{Experiment 2: Settings and Results}

This subsection shows the rescheduling performance. The settings of devices and nodes, and three comparison models are the same as Experiment 1. Each task or node in Experiment 2 has $probability \in \left[10\%, 90\% \right]$ to involve uncertain events, where each task or node can only involve one uncertain event during the simulation. 
\textbf{Table \ref{E2_ue}} shows the settings of uncertain events. We tested the rescheduling in terms of \textit{success rate} and \textit{response time} \cite{nair2022reinforcement}. The success rate refers to the proportion of tasks that are successfully executed after the rescheduling process. The response time refers to the time spent on rescheduling.

\begin{table}[t]
  \caption{Uncertain event settings}
   \setlength{\tabcolsep}{6mm}
  \label{E2_ue}
  \begin{tabular}{ll}\toprule
\textit{Parameters} & \textit{Values}\\
\hline
Task $r$ \& $s$ \& $l$ & increased by  [10\% , 100\%]\\
Task $dl$ & [20, 50] sec in advance\\
Node $ram$ \& $cpu$ & decreased by [20\%, 50\%]\\
Node $bw$ & decreased by 50\%\\
 \bottomrule
  \end{tabular}
\end{table}

\begin{figure}[t!]
\centering
\includegraphics[width=8.5cm]{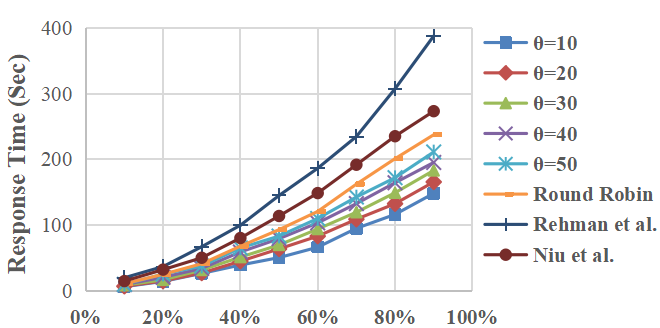}
\caption{Rescheduling response time}
\label{response_time}
\end{figure}

\begin{figure}[t!]
\centering
\includegraphics[width=8.5cm]{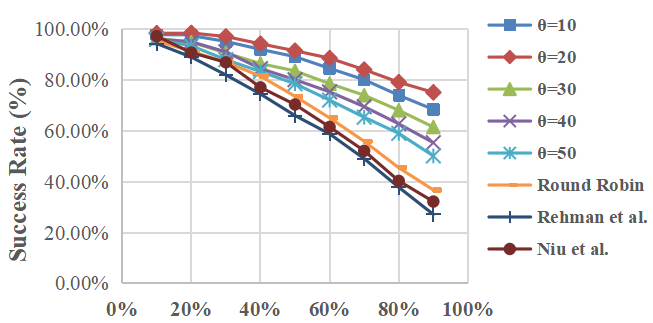}
\caption{Rescheduling success rate}
\label{success_rate}
\end{figure}

\textbf{Figure \ref{response_time}} and \textbf{Figure \ref{success_rate}} show the comparisons between our method and three comparison models. Because three comparison models have no rescheduling mechanism, when uncertain events occur, three models will re-allocate resources for unfinished tasks based on the scheduling algorithm. 
As \textbf{Figure \ref{response_time}} shows, when $\theta$ increases from 10 to 50, the response time increases correspondingly. Because agents solve uncertain events through communications and computations, where the $\theta$ determines the range of agent interactions. As a result, when the response time increases, the task success rate will decrease (shown as \textbf{Figure \ref{success_rate}}). \textbf{Figure \ref{success_rate}} also shows that, the rescheduling owns the highest success rate when $\theta=20$, while $\theta=10$ and $30$ ha similar success rates. By comparing two experimental results, it can be concluded that, when $\theta \in \left[ 10,20 \right]$, the increase of $\theta$ enables more node agents to provide proposals for device agents, so as to increase the rescheduling success rate. When $\theta \geq 20$, this means that device agents start to consider remote nodes. In this case, the data transmission time is increased due to the increase in distances, so the quantity and quality of proposals generated by node agents will not improve. However, the increased response time makes some urgent cannot be completed before deadlines, so as to decrease the success rate. 


Both methods proposed by Niu et al. and the method proposed in this paper can cope with uncertain events by the use of decentralized agents. However, in our method, device agents and node agents try to solve the uncertain events in a local range first, so the impact of uncertain events on task executions is lower than that of Niu et al.'s method.
The Round-Robin and Rehman et al. methods deal with the uncertain events in a centralized manner, where the subsequent occurrence of uncertain events will impact the previous rescheduling process, so the decision-makings are hard to made. Long response time will cause deadline-critical tasks cannot be scheduled in time, so as to decrease the task success rate.

\subsection{Discussion}

Experiment 1 shows that our proposed agent-based scheduling framework can achieve a different performance by adjusting the value of $\theta$. The value of $\theta$ limits the interaction scale of agents during the scheduling process. In this case, each agent will not search for and interact with all other agents. Agents only focus on $\theta$ number of nodes with high priority, where the complexity of each agent is subject to $O(\theta)$, rather than $O(i*p)$, where $i$ denotes the number of device agents and $p$ denotes the number of nodes. 

For the rescheduling purpose, the proposed framework has a quick response mechanism driven by distributed agents. Distributed agents can easily detect uncertain events by themselves and react to events autonomously. Our framework de-centrally solves uncertain events in a small range controlled by $\theta$, where no central or coordinator agents are involved. In this case, executions of other tasks will not be affected. The experimental results also indicate that decentralized agent-based approaches have the potential to minimize the impact of unexpected events during task executions and solve uncertain events with high success rates and short response time.

\section{Related Work} \label{related_work}

Fog computing task scheduling problems have been widely discussed in the literature. Traditional scheduling algorithms, such as the \textit{Round Robin} and \textit{First Come First Served}, and heuristic scheduling algorithms (e.g., min-min and max-min) \cite{rehman2018min}, were considered easy to implement in small and medium configurations \cite{anglano2008scheduling,tychalas2020scheduling}. Traditional algorithms mainly focus on the optimization of scheduling results which centrally allocate resources for tasks. The premise of these algorithms is that the information of all tasks and resources is known in advance. However, entities in fog computing environments are geographically distributed, so it is difficult to collect and respond to real-time information in a centralized manner. Besides, most existing scheduling approaches for fog computing rarely consider task rescheduling problems when uncertain events occur. Since the fog nodes are distributed at the edge of the network, the stability of fog nodes is worse than that of clouds, and the cost of placement and maintenance is expensive, where efficient rescheduling approaches are needed \cite{mahmud2020application}.

The decentralized nature of MAS is in line with that of fog computing. 
Niu et al. \cite{niu2020gmas} proposed a \textit{Geo-aware MAS} that leverages individual agents to cooperate and negotiate with each other to allocate computing resources for IoT tasks. A Geo-aware cost model was presented to decrease the impact of resource locations on traffic costs. Fellir et al. \cite{fellir2020multi} proposed a MAS-based model for task scheduling in the fog computing environment whose objective is to balance the performance of task executions and the utilization of computing resources. 
However, these methods also rarely discuss the task rescheduling problem. To fill this research gap, this paper presents an agent-based task scheduling framework for fog computing and considers to solve four types of typical uncertain events. 

\section{Conclusion}\label{conclusion}

This paper proposed an agent-based task scheduling framework for fog computing under the consideration of four types of uncertain events. Experimental results demonstrated that our framework could flexibly schedule IoT tasks in fog computing environments, and handle uncertain events with a higher success rate and lower response time than that of comparison methods. Further work will extend the current framework to handle more types of uncertain events with more real-world constraints.

\bibliographystyle{ACM-Reference-Format} 
\bibliography{aamas_paper}


\end{document}